\documentclass[12pt,preprint]{aastex}

\usepackage{graphicx,apjfonts,emulateapj5}

\shorttitle{Counterclockwise magnetic fields in the Norma spiral arm}
\shortauthors{Han et al.}

\begin{document}
\title{Counterclockwise magnetic fields in the Norma spiral arm}
\author{J. L. Han\altaffilmark{1,6},
        R. N. Manchester\altaffilmark{2},
	A. G. Lyne\altaffilmark{3},
	and
        G. J. Qiao\altaffilmark{4,5,6}
	}

\altaffiltext{1}{National Astronomical Observatories, Chinese 
	Academy of Sciences, 20A Datun Road, Chaoyang, Beijing 100012, China.
	Email: hjl@bao.ac.cn}
\altaffiltext{2}{Australia Telescope National Facility, CSIRO, PO Box 76,
        Epping, NSW 1710, Australia. Email: Dick.Manchester@csiro.au}
\altaffiltext{3}{University of Manchester, Jodrell Bank Observatory,
	Macclesfield, SK11 9DL, UK. Email: agl@jb.man.ac.uk}
\altaffiltext{4}{CCAT (World Laboratory), PO Box 8730, Beijing 100080, China.}
\altaffiltext{5}{Department of Astronomy, Peking University (PKU),
	Beijing 100871, China. Email: gjq@bac.pku.edu.cn}
\altaffiltext{6}{Beijing Astrophysical Center, CAS-PKU, Beijing 100871, China.}

\begin{abstract} 
Pulsars provide unique probes of the large-scale interstellar magnetic
field in the Galactic disk. Up to now, the limited
Galactic distribution of the known pulsar population has restricted these
investigations to within a few kiloparsec of the Sun. The Parkes multibeam
pulsar survey has discovered many more-distant pulsars which enables us
for the first time to explore the magnetic field in most of the nearby
half of the Galactic disk.  Here we report the detection of counterclockwise
magnetic fields in the Norma spiral arm using pulsar rotation measures.
The fields are coherent in direction over a linear scale of $\sim 5$
kpc along the arm and have a strength of $-4.4\pm0.9$ $\mu$G.  The
magnetic field between the Carina-Sagittarius and Crux-Scutum arms
is confirmed to be coherent from $l\sim45\degr$ to $l\sim305\degr$
over a length of $\sim 10$ kpc. These results strengthen arguments for
a bisymmetric spiral model for the field configuration in the Galactic
disk. 
\end{abstract}

\section{Introduction}
The origin of galactic magnetic fields is a long-debated issue,
which relies heavily on good observational descriptions of the
structure and strength of magnetic fields in a galaxy \citep{zh97,kro94}.
The current magnetic field configuration could be the result of a
frozen-in primordial field in a protogalactic gas cloud which
was strengthened during the collapse and formation of a galaxy
\cite[e.g.,][]{kul90}. The strength of the primordial field should
be $\sim10^{-9}$~G to give the typical field strength (a few $\mu$G)
presently seen in galaxies.  A field of primordial origin would have
many reversals of field direction in the galactic disk. However, the
presently favored model for the field origin is dynamo amplification
of a seed field by the inductive effects of the fluid motions of the
interstellar medium \citep[e.g.,][]{rss88}. Seed fields of at least
$10^{-13}$~G are required.  Computer simulations of dynamo action in
galaxies, although far from mature, show a variety of field structures
depending on the initial conditions and the details of the dynamics
of medium. Usually the fastest-growing mode is axisymmetric in
configuration \citep{fs00}.

In the last two decades, polarization observations of synchrotron
emission at centimeter wavelengths have revealed that in nearby galaxies
the intrinsic magnetic field ``vectors'', i.e. {\bf E}+90$\degr\leftrightarrow$
{\bf B} with Faraday rotation corrected, are impressively aligned along
the optical arms \citep{bbm+96, bec00}. Such polarization can arise
not only from the regular large-scale magnetic fields as conventionally
believed but also from an anisotropy in random fields \citep{lai02},
for example, compressed in one dimension by a density wave shock in
spiral galaxies. Evidence for coherence of field {\it directions},
that is, including the sense of the field, over large scales in the
disks of external galaxies is fairly weak at present \citep[e.g.][]{hbe+99}.
In our Galaxy, however, rotation measures (RMs) for pulsars and extragalactic
radio sources have revealed large-scale ordering of magnetic field
directions in the Galactic disk, with several reversals between or
within the spiral arms \citep[e.g.][]{sk80,rl94,hmq99}. Determination
of the form of such directional coherence is crucially important for
studies of the origin of galactic magnetic fields. Field reversals
could be preserved from seed fields or there could be a mixture of
dynamo modes generating the field.

Pulsar RM data suggest that the pitch angle of the local field is
about $-8\degr$ and its strength is about 1.5 $\mu$G \citep{han01}.
Between the Perseus and the Carina-Sagittarius arms the field
direction is clockwise if viewed from the North Galactic
pole\footnote{We use this convention throughout this Letter}, as
illustated by thick arrows in Fig.~\ref{fg:rm-xy}. In the outer
Galaxy, \citet{ls89}, \citet{ccsk92} and \citet{hmq99} suggested that
the direction of the regular field reverses exterior to the Perseus
arm, while Canadian groups argued for no reversal between the local
arm and the Perseus arm \citep{bt01} or in the outer Galaxy
\citep{val96}.  In the inner Galaxy, RMs of pulsars and extragalactic
sources suggested reversed or counter-clockwise fields between the
Carina-Sagittarius arm and the Crux-Scutum arm \citep{tn80, sk80,
rk89}, later confirmed by new pulsar observations \citep{rl94}. Magnetic
fields beyond the Crux-Scutum arm but closer than the Norma arm {\it
may} have a clockwise direction, as indicated by the negative RMs
around $l\sim20\degr$ at about 5kpc \citep{rl94} and an RM increase at
about the same distance around $l\sim325\degr$.

It would be intriguing to know if the regular local fields are parts
of a larger scale {\it coherent} field structure and if the regular
fields in distant parts of the Galactic disk show reversals.  Since we
sit within the disk, it is not possible to have a birds-eye view of
the polarized emission of our Galaxy, as for suitably oriented nearby galaxies
\citep{bbm+96}.  Analyses of starlight and Galactic background
polarization data show magnetic properties only relatively locally to
the Sun \citep[e.g.][]{hei96a, flpt02} or the transverse field in
polarized features \citep{dhjs97, ufr+99}. Zeeman splitting of OH
(or other) masers yields the line-of-sight direction and magnitude of
the magnetic fields only within star-forming regions \citep{rs90,
cv95}. RMs of extragalactic radio sources are integrated along the
line of sight through the entire Galactic disk, and are always affected
to some extent by Faraday rotation local to the sources\footnote{The RM
arising within extragalactic radio sources can be more than 100 rad~m$^{-2}$,
as shown by the RMs of sources \citep{skb81} near the two Galactic
poles ($|b|>70\degr$) where the Galactic RM contribution is only
$\sim (+-) 3$ rad~m$^{-2}$ \citep[see Sect.3.3 in][]{hmq99}.}. Pulsars
do not seem to have such intrinisic RM contributions. Moreover,
pulsars cover a range of distances and therefore offer the best
opportunity to investigate the 3-dimensional structure of the
interstellar magnetic field over a substantial fraction of the
Galactic disk.

Recently, the Parkes multibeam pulsar survey \citep{mlc+01} has
discovered more than 600 pulsars, many of which are widely distributed over
more than half of the Galactic disk. This sample therefore provides a
unique opportunity for investigation of the magnetic field structure
in the inner Galaxy. We have observed about 240 southern pulsars with
the Parkes 64-m telescope of the Australia Telescope National Facility
and obtained 202 new and improved RMs. Based on these
data, we report the first firm detection of counterclockwise
magnetic fields in the Norma spiral arm. A more comprehensive analysis
of the full data-set will be published in a future paper.

\section{Pulsar observations and rotation measures}

Pulsar polarization observations were made in two observation
sessions, 1999 December 12 -- 17 and 2000 December 14 -- 19, using the
Parkes 64-m telescope with the central beam of the multibeam receiver
\citep{swb+96} which is sensitive to orthogonal linear polarizations.
Bands centered on 1318.5 MHz with a bandwidth of 128 MHz were
processed in the Caltech correlator \citep{nav94}, which gives 128
lags in each of four polarization channels and folds the data
synchronously with the pulsar period with up to 1024 bins per pulsar
period. The data were transformed to the frequency domain, calibrated
and dedispersed to form between 8 and 64 frequency sub-bands, with
the number depending on the pulsar dispersion measure (DM), in each
of the four Stokes parameters ($I,Q,U,V$), and corrected for variations in
parallactic angle and ionospheric Faraday rotation. Because of reduced
gain near the band edges, about a quarter of the bandwidth
(mostly at the lower end) was abandoned, making an effective bandwidth
of about 90 MHz. To reduce residual instrumental effects such as feed
cross-coupling, observations of each pulsar were made in pairs with
the receiver rotated to orthogonal feed angles of $-45\degr$ and $+45\degr$.

In off-line analysis, we summed the sub-band data over a set of
trial RMs, searching for a peak in the linearly polarized intensity
$L = (Q^2 + U^2)^{1/2}$.  Normally, a range of at least 
$\pm$2000 rad~m$^{-2}$ was searched with a step of about 20 rad~m$^{-2}$. 
The sub-band data were then summed with the RM at which the most
significant peak was found, to form two sets of Stokes profiles for
the upper and lower halves of the available bandpass. A final value
for the RM and its uncertainty were then determined from weighted
position-angle differences across the pulse profiles. 
We obtained a total of 202 RMs in two
sessions. Comparing the RM values of 11 pulsars with independent
measurements by \citet{cmh91} and \citet{vdhm97}, we found that our
results are consistent with and generally have better precision than
previous values. 

\begin{figure*} 
\includegraphics[width=150mm,angle=270]{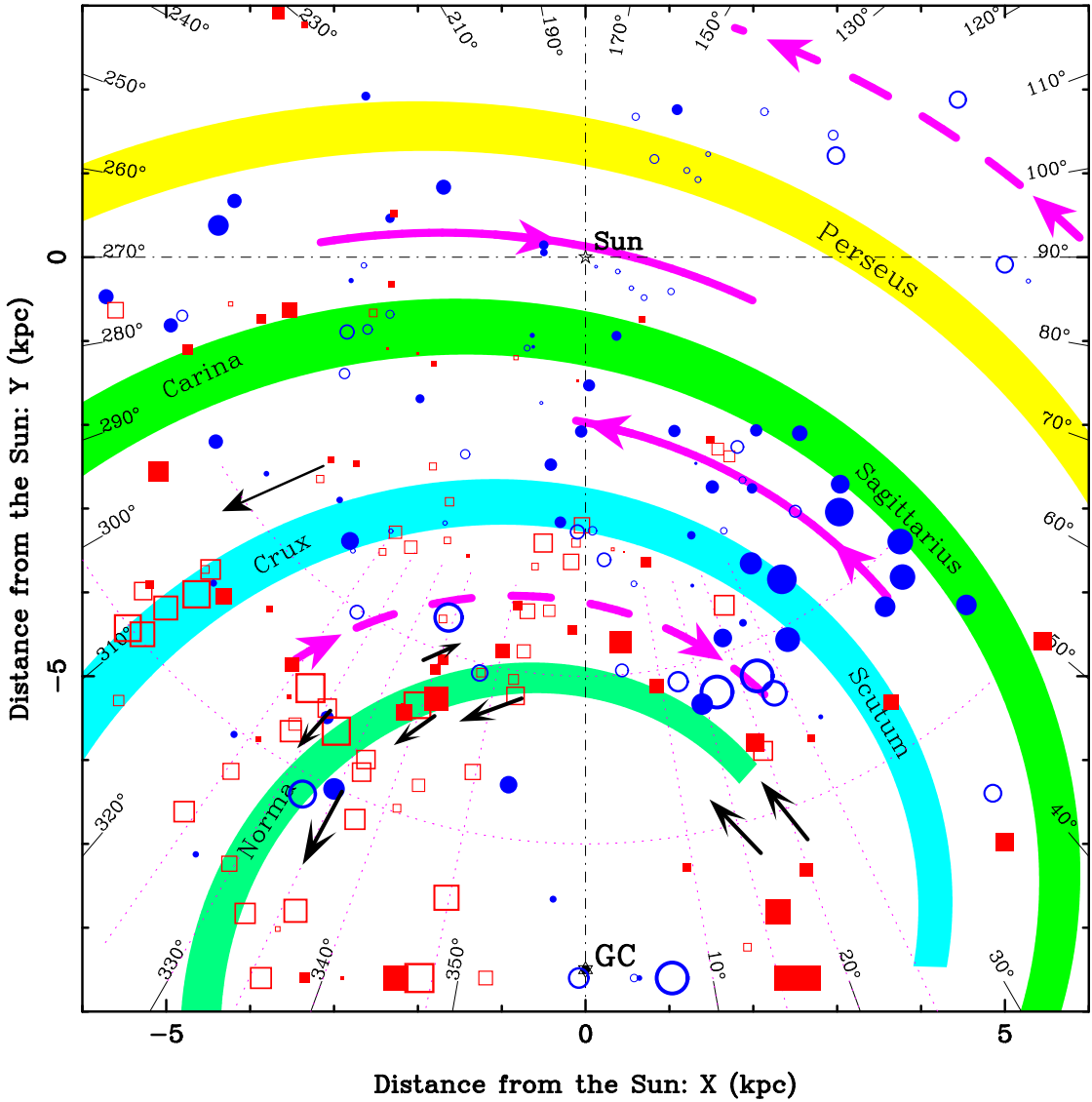}
\caption{Rotation measures of pulsars having $|b|<8\degr$, projected
on to the Galactic Plane. The areas of the symbols are proportional
to the RM magnitudes, with limits of 5 and 800
rad m$^{-2}$. Filled symbols represent positive RMs and open symbols
indicate negative RMs. New measurements are indicated by squares.
Pulsar distances were calculated by using the model of \protect\citet{tc93}.  
Approximate locations of four spiral arms are indicated
\protect\citep{gg76,dwbw80,ch87a}.  The generally accepted magnetic
field directions between the Perseus and Carina-Sagittarius arms and
between Carina-Sagittarius and Crux-Scutum arms are illustrated by
thick lines and arrows. Evidence exists for field directions exterior
to the Perseus arm and between the Crux-Scutum and Norma arms, as
shown by thick dashed lines and arrows (see text), but more data are
needed to confirm them. Field directions near the Norma arm are
plotted according to fitted pulsar RM gradients as shown in Fig.\ref{fg:rm-dm}.
Dotted lines indicate distances of 5 and 7 kpc from the Sun and some
Galactic longitudes of interest.}
\label{fg:rm-xy}
\end{figure*}

To study the magnetic field in the Galactic disk we consider only
pulsars with $|b|<8\degr$. There are now 357 such pulsars with measured
RMs, 170 of which are new measurements. The new negative RMs at direction
around $l\sim310\degr$ and distance of 7 kpc confirm the continuity of
counterclockwise fields between the Carina-Sagittarius and Crux-Scutum
arms, which was originally suggested by the dominant positive RMs in
the area between the two arms in the first quadrant (see
Fig.~\ref{fg:rm-xy}). The fields are therefore coherent over a scale
of 10 kpc along the arm. The RM data in the region $325\degr<l<25\degr$,
together with previously published values, enable us to identify clearly
and to measure the regular magnetic field in the vicinity of the Norma
arm, which we will discuss in detail below.

The mean line-of-sight component of the magnetic field (in $\mu$G)
along the path to the pulsar, weighted by the local electron density,
is given by 
\begin{equation}
\langle B_{||}\rangle = 1.232\; {\rm RM}/{\rm DM} 
\end{equation}
where RM is in units of rad~m$^{-2}$ and the DM is in units of
cm$^{-3}$ pc \citep{man74}. The mean field strength between any two
points at distances of $d0$ and $d1$ in a given direction can be
found from the {\it gradient} of the RM 
\begin{equation}
\langle B_{||}\rangle_{d1-d0}  = 1.232\;\frac{\Delta {\rm RM}}{\Delta
{\rm DM}}
\label{eq:rmgrad}
\end{equation} 
where $\Delta{\rm RM} = {\rm RM}_{d1} - {\rm RM}_{d0}$ and $\Delta{\rm
DM} = {\rm DM}_{d1} - {\rm DM}_{d0}$. Of course, in practice, it is not
possible to have sources at different distances along exactly the same
line of sight, so that sources within a small area on the sky, typically
a few degrees across, are taken to represent the column in that direction. 

This method of using RM gradients was used first by \citet{ls89} and
\citet{ccsk92} to show the field reversals near the Perseus arm using
RM data for pulsars and extragalactic radio sources. Later, \citet{rl94}
and \citet{hmq99} used it to show the field reversal near the
Carina-Sagittarius arm and near the Crux-Scutum arm. The RM effects of
local bubbles which are tens of degrees across \citep[see][]{val84}
have no effect on the differential analysis for more distant regions. 

\begin{figure*} 
\includegraphics[height=160mm,angle=270]{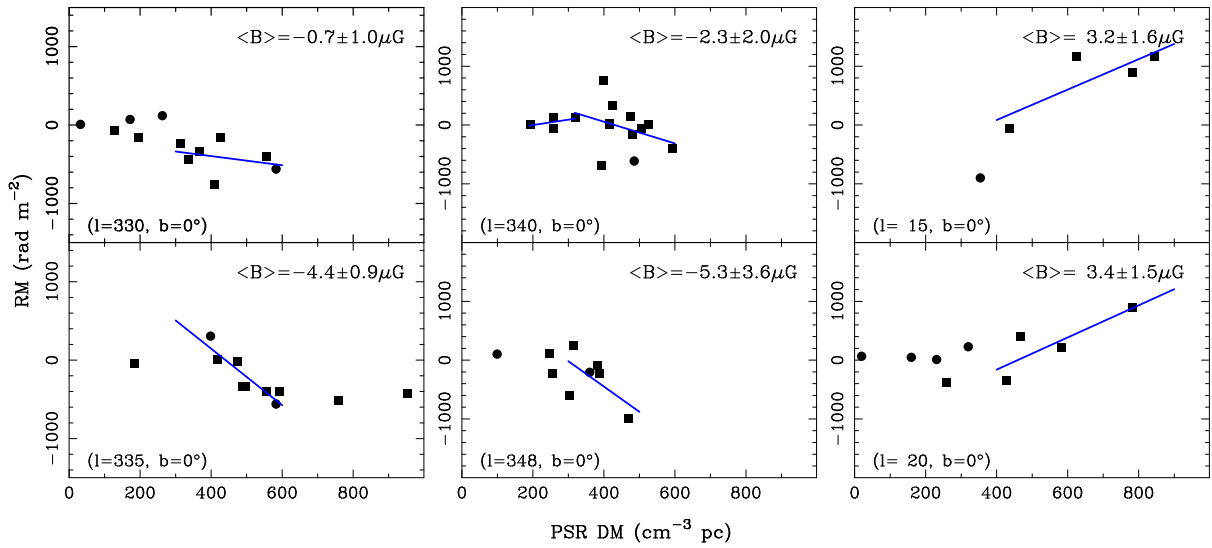}
\caption{Variations of RM versus DM for pulsars within $3\degr$ of six
lines of sight passing near to or across the Norma arm. New RM
measurements are indicated by squares. Straight-line fits to the data
near the Norma arm are indicated by the lines in each panel. Toward
$l=340\degr$, we also fit closer pulsars giving information on the
field between the Crux and Norma arms.}\label{fg:rm-dm}
\end{figure*}

In Fig.~\ref{fg:rm-dm} we plot the RM against DM for pulsars having RM
uncertainties of $<50$ rad~m$^{-2}$ and lying within $3\degr$ of six
lines of sight which cross or are tangential to the Norma spiral
arm. It is very clear that for DMs greater than $\sim 300$ cm$^{-3}$
pc, the pulsar RMs are systematically decreasing in the longitude
range of $l\sim 325\degr$ to $350\degr$ and increasing in the range of
$l\sim 10\degr$ to $25\degr$. We have used both the more detailed
electron density model of \citet{tc93} and the simple axisymmetric
model of \citet{gbc01} to estimate the pulsar distances from
dispersion measure values. Both models suggest that these pulsars are
either associated with or beyond the Norma arm, although with a
distance uncertainty of about 20\% in general. Since the simple model
is almost unconstrained interior to a Galactocentric radius ($R$) of
4 kpc, we have plotted pulsars in Fig.~\ref{fg:rm-xy} using distances
given by the model of \citet{tc93}.

The strength of the regular fields can be estimated directly from the
gradient of the RM versus DM variations (Equation~\ref{eq:rmgrad}).
We fitted straight lines using a least-squares procedure to the RM
versus DM data for pulsars with 300 cm$^{-3}$ pc $<$ DM $<$ 700
cm$^{-3}$ pc in the longitude range of $l\sim 325\degr$ to $350\degr$,
and DM $>400$ cm$^{-3}$ pc in the range of $l\sim 10\degr$ to
$25\degr$. These DM ranges correspond to pulsars which lie in the
vicinity of the Norma arm. In directions toward $l=340\degr$, pulsars
with DM in the range 200 -- 300 cm$^{-3}$ pc were also fitted, giving
information on the field between the Crux and Norma arms. The results
are shown Fig.~\ref{fg:rm-dm}. Pulsars around $l\sim335\degr$ show the
strongest evidence for a systematic field within or inside the Norma
arm with an average field strength of $-4.4\pm0.9\mu$G.  Data in other
panels of Fig.~\ref{fg:rm-dm} support or are consistent with this
result. These fits are also shown as vectors in Fig.~\ref{fg:rm-xy},
where an azimuthal configuration for the field has been assumed.

The derived field strength is two or three times greater than the
local regular fields \citep{hq94, id98}.  Since the `vertical' and
`radial' components of the large-scale field are about an order of
magnitude weaker that the azimuthal component \citep{hq94,hmq99}, this
is good evidence for a large-scale counterclockwise magnetic field in
and interior to the Norma arm, with directional coherence over at least
5 kpc along the arm. Based on the RM data in the Galactic longitude
range $320\degr<l<335\degr$, this counterclockwise field may extend
over more than 2 kpc in Galactocentric radius, roughly in the range 4
-- 6 kpc.

\section{Discussion}
The directions of large-scale regular fields in the Norma region shown
in Fig.~\ref{fg:rm-xy} clearly reveal for the first time the
counter-clockwise large-scale field within and interior to the Norma
spiral arm.  Extragalactic radio sources between longitudes of
$325\degr$ to $330\degr$ have large negative RMs \citep{gdm+01},
consistent with a regular magnetic field parallel to the line of sight
near the arm tangent.

Although evidence for a clockwise field between the Crux-Scutum
arm and the Norma arm is relatively weak at present, the presence of a
counterclockwise field in the Norma arm indicates another reversal in
the directions of large-scale azimuthal fields in the Galactic disk.
Further observations in the longitude ranges $310\degr-325\degr$
and $ 25\degr- 40\degr$ are needed to confirm the existence and extent
of the clockwise field between the Crux-Scutum and Norma arms. If this
is confirmed, there will be good evidence for four reversals in the
azimuthal field. Present indications are that these occur within the
spiral arms, although this is not yet clear.

These field reversals are consistent with the bisymmetric spiral model
for the Galactic disk field \citep{sk80, sf83, hq94, id98}. They appear
inconsistent with the axisymmetric model of \citet{val96}, since this
model does not allow any field reversal exterior to the Perseus arm or
interior to the Crux-Scutum arm.  The field reversals are most probably
a remnant of primordial fields at the time our Galaxy formed. These act
as `seed' fields for dynamos operating in the conductive interstellar
medium \citep{kul99}. Dynamo action is necessary to counteract diffusion
of the large-scale field structure. However, we cannot exclude the
possibility that such a bisymmetric field could be the result of a
strong non-axisymmetric configuration of the dynamo induced, for example,
by tidal interaction with the Magellanic Clouds \citep[see][]{mos96}. 

It is desirable to understand why our Galaxy has several reversals in
the thin Galactic disk as revealed by pulsar and extragalactic RMs,
while few reversals are known in nearby galaxies. Reversals are known
to exist in M51 \citep{bhk+97}, in M81 \citep{kbh89} and possibly in
NGC2997 \citep{hbe+99}. Polarization observations of external galaxies
alone cannot distinguish reversals in the field direction.  RM maps
deduced from multifrequency polarization maps generally have a lower
resolution than interarm separations or field-reversal scales, and
hence it is difficult to identify reversed fields \citep[see also][]{hei95}.
The polarized emission observed in nearby galaxies mainly comes from a
thin disk and it indicates the orientation of regular fields or possibly
the anisotropy of random fields in the disk. However, the Faraday
rotation of the polarized emission, which has been used to infer the
presence or absence of field reversals in nearby galaxies, may
originate mainly in the intervening thick disk or halo. Even if the
thin disk field has reversals, the halo field may be axisymmetric
and have no reversals. For example, there is evidence that our Galaxy
has an axisymmetric halo field generated by an A0 dynamo \citep{hmbb97}. 
The mixture of axisymmetric and bisymmetric spiral fields in external
galaxies \citep{bec00} may be explained if the thin disk, thick disk
and halo are together responsible for the RM distribution.

The coherence of field directions in our Galaxy revealed by pulsar RMs
independently show that the regular magnetic fields do indeed exist on
galactic scales. Though we can not exclude that some of the linear
polarization is caused by anisotropic random fields, our results show
that at least some of the linear polarization arises from large-scale
coherent fields.  Therefore, it is likely that polarization ``vectors''
observed in the external galaxies \citep{bbm+96, bec00} represent the
large-scale regular magnetic field.

An increase in strength of the large-scale field with decreasing
Galactocentric radius is expected in both dynamo models \citep{rss88}
and for fields of primordial origin \citep{kul90}. With a few more
measurements, it will be possible to determine if the field strength
obeys the relation $B \propto 1/R$, which is a critical profile for
the field to affect the dynamics of the Galaxy \citep{bf00, kro94}.

\section*{Acknowledgments}
We thank the referee for helpful comments and Rainer Beck and Richard
Wielebinski for useful discussions, all of which led to improvements in
the paper.
JLH thanks the exchange program between CAS and CSIRO for support of
the visits at the Australia Telescope National Facility in 1999 and 2000.
His research in China is supported by the National Natural Science
Foundation (NNSF) of China (19903003 and 10025313) and the National
Key Basic Research Science Foundation of China (G19990754) as well
as the partner group of MPIfR at NAOC.
GJQ thanks the NNSF of China for support for his visits at ATNF.
The Parkes radio telescope is part of the Australia Telescope which is
funded by the Commonwealth Government for operation as a National
Facility managed by CSIRO.


\end{document}